\documentclass[seceq,supplement]{ptptex}

\usepackage{graphicx}
\usepackage{wrapft}

\def\lsim{\mathrel{\rlap{\lower 4pt \hbox{\hskip 1pt $\sim$}}\raise 1pt
\hbox {$<$}}} 
\def\gsim{\mathrel{\rlap{\lower 4pt \hbox{\hskip 1pt $\sim$}}\raise 1pt
\hbox {$>$}}}

\newcommand{\msun}{M_{\odot}}

\title{
Jets and Black Holes in Hypernova Explosions%
}

\author{
Keiichi \textsc{Maeda}\footnote{E-mail: maeda@astron.s.u-tokyo.ac.jp } 
and Ken'ichi \textsc{Nomoto}%
}

\inst{
Department of Astronomy, University of Tokyo, Tokyo 113-0033, Japan 
}

\abst{
Bipolar explosion models for hypernovae (very energetic supernovae), 
associated with the black hole formation, are presented. 
The model features are as follows: 
(1) Fe-peak elements are ejected at high velocities to reach the surface region, 
while dense O-rich materials occupy the central region at low velocities. 
(2) The abundance ratios (Zn, Co)/Fe are enhanced while the ratios (Mn, Cr)/Fe 
are suppressed. 
The overall abundances are consistent with those observed in extremely metal-poor stars 
formed at the very early phase of the Galaxy, suggesting significant contribution 
of such explosions to the early Galactic chemical evolution. 
(3) Certain relations among the optical luminosity, the chemical composition, 
and the property of the central remnant are predicted. 
Such relations may serve for identifying possible gravitational wave emitters 
in the future. 
}

\begin{document}

\maketitle

\section{Introduction}

Discovery of a new class of supernovae (SNe), called 'hypernovae', 
which have distinctly large kinetic energies ($E_{51} = E/10^{51}$ergs $\gsim 10$) 
compared with previously known supernovae ($E_{51} \sim 1$) 
is one of the most exciting topics in recent 
studies on supernovae. For a review of hypernovae, 
see Refs.~\citen{nomoto2003a,nomoto2003b}. 

Modeling spectra and light curves of these objects has uncovered an interesting relation 
between progenitor's masses ($M_{\rm MS}$) and outcomes ($E$, $M$($^{56}$Ni)). 
(See Ref.~\citen{nomoto2003b} and references therein.)
There is an apparent transition around $20-25\msun$ in these properties, 
from normal SNe to two branches of hypernovae and faint SNe. 
The modeling of hypernovae suggests that 
$E$ and $M$($^{56}$Ni) are larger for larger $M_{\rm MS}$ in this 'hypernova branch'. 
The transition implies that the explosion mechanism of stars with $M_{\rm MS} \gsim 20-25\msun$ 
might be different from that of stars with $M_{\rm MS} \lsim 20 \msun$, 
especially for hypernovae given their large energies. 
(For faint SNe, the small $E$ may simply be attributed to a large gravitational 
binding energy.) 

In this paper, we present bipolar explosion models for hypernovae. 
We describe the hydrodynamic and nucleosynthetic results, 
and predict how the nucleosynthetic yield, 
the optical luminosity, and the property of the central remnant are related.

\section{Hydrodynamics}

\begin{table}
\caption{Models and Results. Masses are in solar mass unit ($\msun$), 
and $\theta_{\rm jet}$ is in degree.
Abundance ratios are normalized at the solar values, i.e., 
[X/Y] $\equiv \log10 ({\rm X/Y}) - \log10({\rm X/Y})_{\odot}$.}
\begin{center}
\begin{tabular}{cccccccccc}
\hline \hline
{Model} & {$M_{\rm MS}$} &  {$M_{\rm REM0}$} & {$\alpha$} &
{$\theta_{\rm jet}$} & {$E_{\rm total}$} &
{$M_{\rm REM}$} & {$M$($^{56}$Ni)} & [S/Si] & [C/O] \\
\hline
40A & 40 & 1.5 & 0.01 & 15 & 10.9 & 5.9 & 1.07E-1 & -0.46 & -1.3\\
40B & 40 & 1.5 & 0.01 & 45 & 1.2 & 6.8 & 8.11E-2  & -0.54 & -1.2\\
40C & 40 & 1.5 & 0.05 & 15 & 32.4 & 2.9 & 2.40E-1 & -0.30 & -1.3\\
40D & 40 & 3.0 & 0.01 & 15 & 8.5 &10.5 & 6.28E-8  & -1.1  & -1.0\\
\hline
25A & 25 & 1.0 & 0.01 & 15 & 6.7 & 1.9 & 7.81E-2  & -0.28 & -0.80\\
25B & 25 & 1.0 & 0.01 & 45 & 0.6 & 1.5 & 1.51E-1  & -0.26 & -0.82\\
\hline
\end{tabular}
\end{center}
\end{table}

\begin{figure}[t]
	\begin{center}
		\begin{minipage}[t]{0.45\textwidth}
			\includegraphics[width=0.925\textwidth]{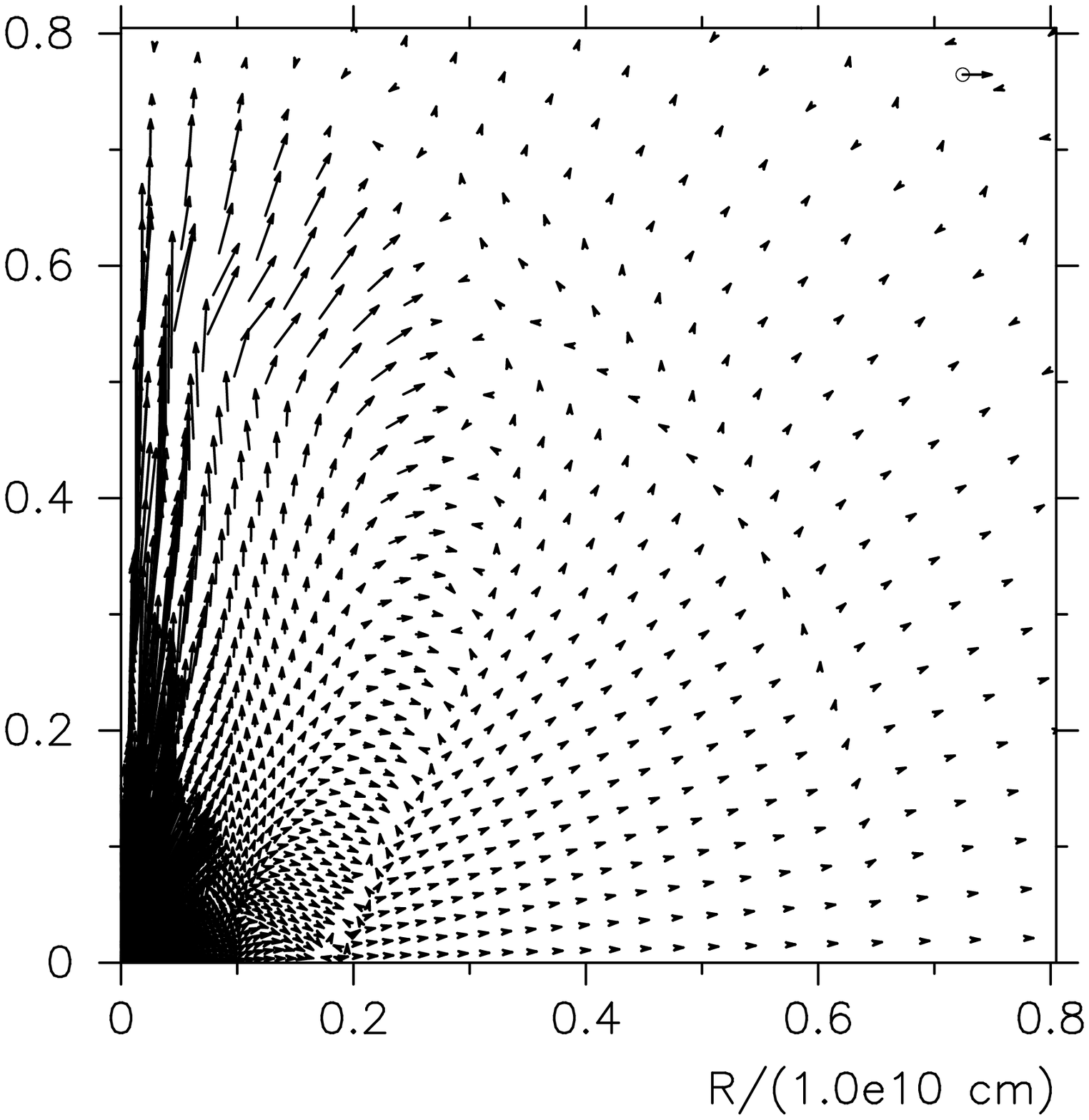}
		\end{minipage}
		\hspace{1cm}
		\begin{minipage}[t]{0.45\textwidth}
			\includegraphics[width=0.925\textwidth]{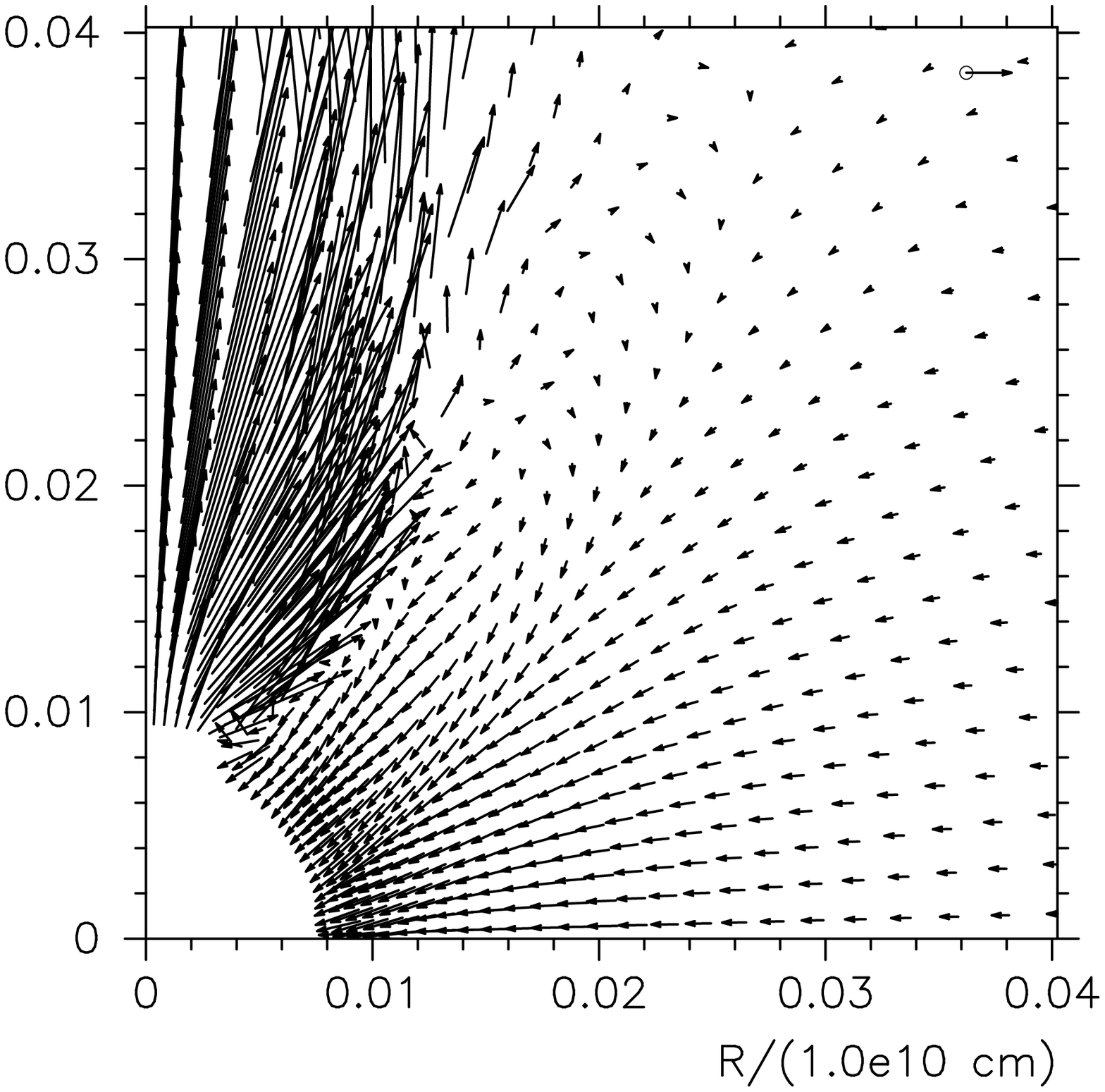}
		\end{minipage}
		\vspace{1cm}
	\end{center}
\caption{Velocity distribution of Model 40A at 1.5 second after the initiation of the jets. 
The right panel shows that in the central region on an expanded scale. 
The reference arrow at the upper right represents $2 \times 10^9$cm s$^{-1}$. 
\label{fig:fig1}}
\end{figure}

The main ingredient of our models is a pair of jets propagating through a stellar mantle. 
At the beginning of each calculation, the central part ($M_r \leq M_{\rm REM0}$) 
of a progenitor star 
is displaced by a compact remnant. The jets are injected at the interface 
with the opening half-angle being $\theta_{\rm jet}$. 
The energy injected by the jets is assumed to be proportional to 
the accretion rate (i.e., $\dot E_{\rm jet} = \alpha \dot M_{\rm accretion} c^2$). 
The models are summarized in Table 1.
The outcome of the explosion depends on the interaction between the jets 
and the stellar mantle, and on the accretion rate 
which is affected by the interaction itself. We follow the hydrodynamic evolution 
and nucleosynthesis in two dimensions. 

Figure 1 shows a snapshot of the velocity distribution of Model 40A 
at 1.5 seconds after the initiation of the jets. 
The jets propagate through the stellar mantle, depositing their energies 
into ambient matter at the working surface. 
The bow shock expands laterally to push the stellar mantle sideways, 
which reduces the accretion rate. 
The strong outflow occurs along the $z$-axis (the jet direction), 
while matter accretes from the side. 
As the accretion rate decreases, the jets are turned off. 
Then the inflow along the $r$-axis turns to the weak outflow. 

The outcome is a highly aspherical explosion. 
The accretion forms a central dense core. 
Densities near the center become much higher than those in spherical models. 
This feature is indeed what is suggested by the spectroscopic \cite{mazzali2000} 
and light curve \cite{nakamura2001,maeda2003a} modeling of hypernovae. 

Other hydrodynamic properties are as follows (Table 1): 
(1) A more massive star makes a more energetic explosion. 
The reason is that a more massive star has a stronger gravity to 
make the accretion rate higher. 
This is consistent with the relation seen in the hypernova branch. \cite{nomoto2003b}
(2) A more massive star forms a more massive compact remnant. 
The remnant's mass increases as the accretion feeds it. 
The final mass $M_{\rm REM}$ reaches typically $>5\msun$ for a $40\msun$ star, 
and $\sim 2\msun$ for a $20\msun$ star. 
The bipolar models provide the way of explosions with black hole formation in a consistent manner. 
Given the discovery of the evidence of a hypernova explosion that accompanied formation 
of a black hole of $\sim 5\msun$ (X-ray Nova Sco; Refs~\citen{israelian1999,podsiadlowski2002}), 
it offers an interesting possibility.

\section{Nucleosynthesis}

\begin{figure}[t]
	\begin{center}
		\begin{minipage}[t]{0.45\textwidth}
			\includegraphics[width=1.0\textwidth]{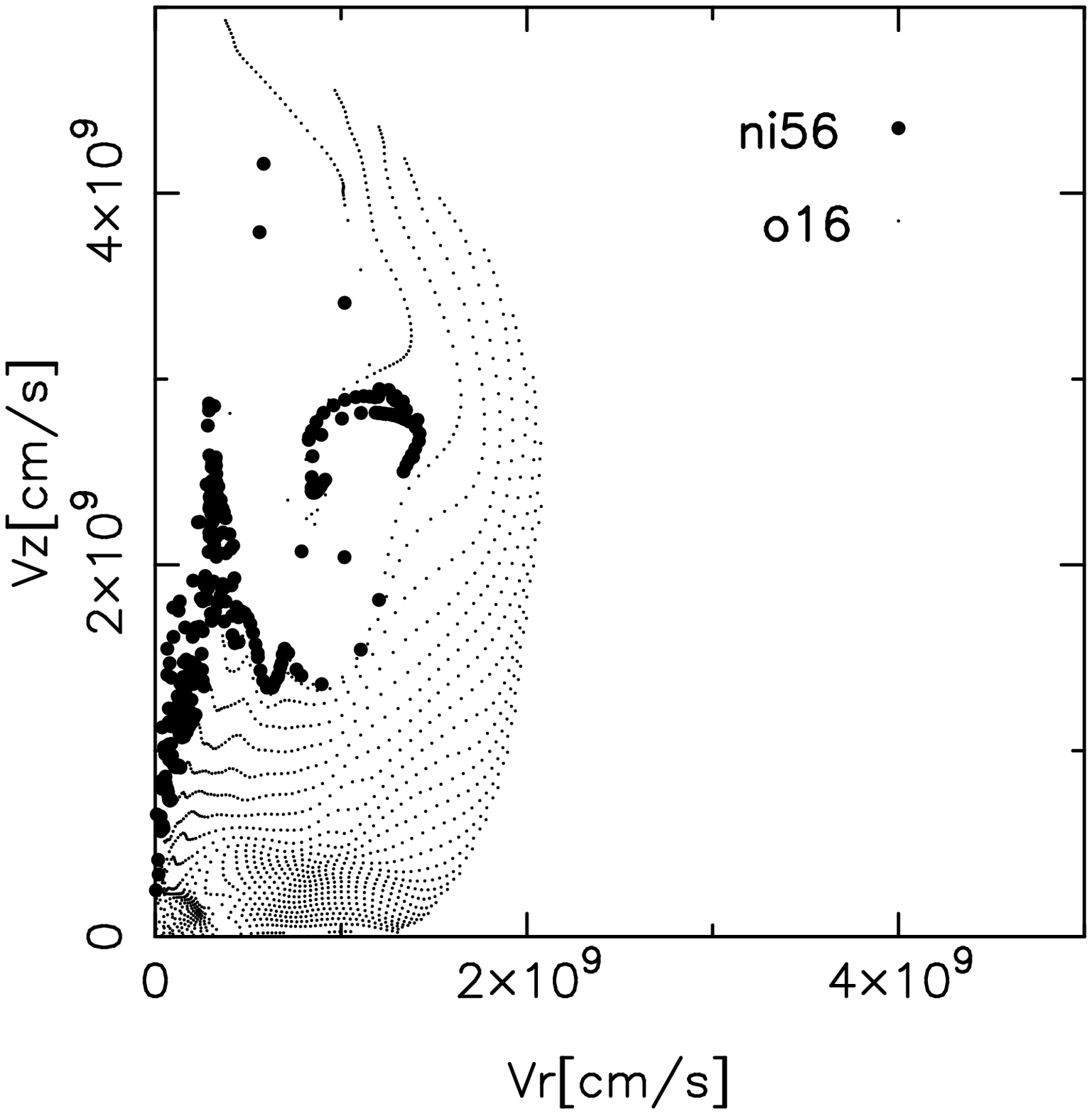}
		\end{minipage}
		\hspace{1.0cm}
		\begin{minipage}[t]{0.45\textwidth}
			\includegraphics[width=1.0\textwidth]{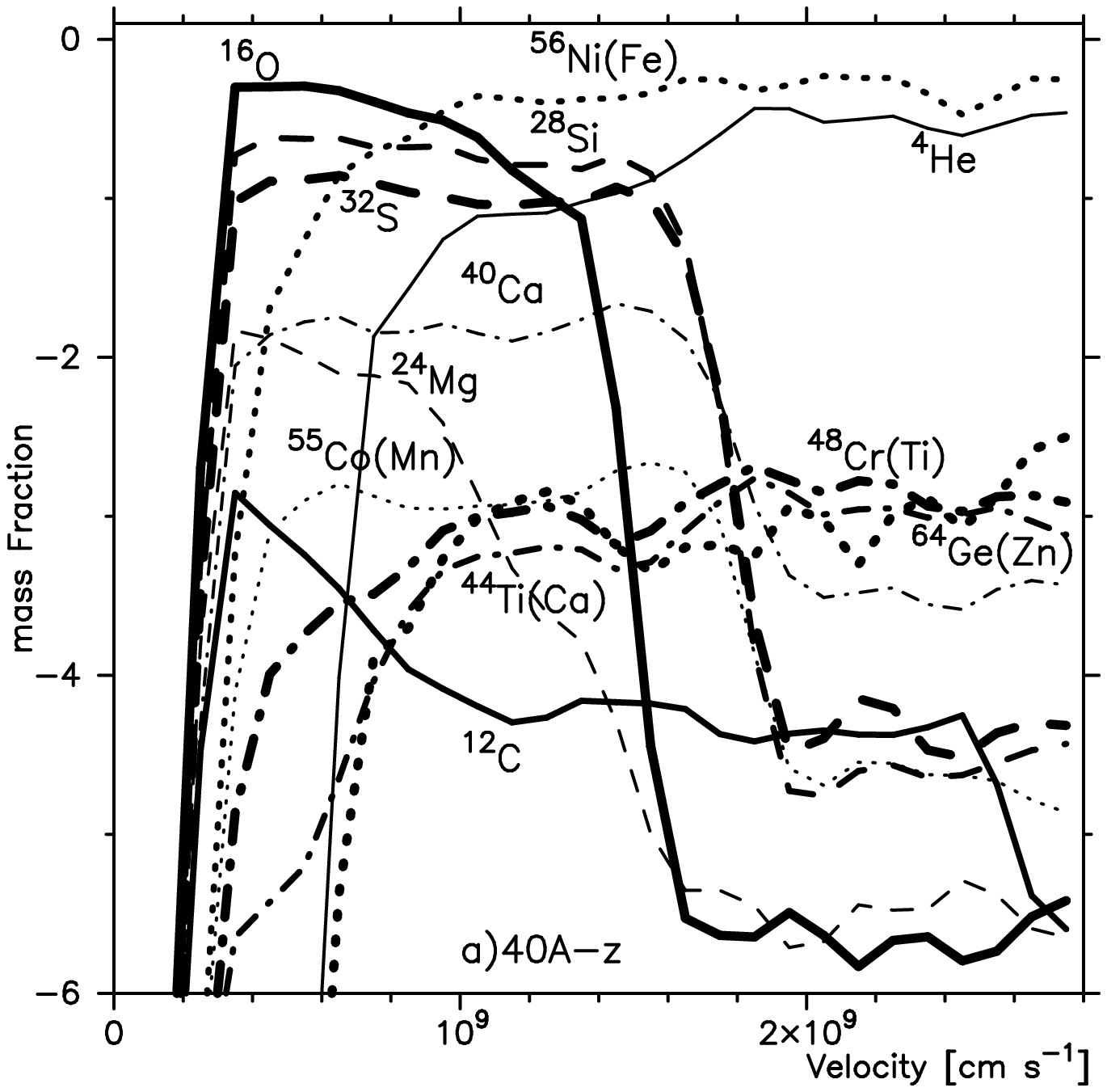}
		\end{minipage}
	\end{center}
	\caption{Left: Distributions of $^{56}$Ni (which decays into $^{56}$Fe: filled circles) and $^{16}$O (dots).  
The mass elements in which the mass fraction of each isotope exceeds 0.1 are plotted. 
Right: Mass fractions of selected isotopes in the velocity space along the $z$-axis of Model 40A. 
\label{fig:fig2}}
\end{figure}

Relatively high temperatures along the $z$-axis and low temperatures along the $r$-axis 
have significant effects on nucleosynthesis. 
It results in highly aspherical distribution of nucleosynthetic products 
as shown in Figure 2. 
The distribution of $^{56}$Ni (which decays into $^{56}$Fe) is 
elongated along the $z$-axis. Such a configuration, i.e., 
high velocity Fe and low velocity O, has been suggested to be responsible 
for the feature in the late phase spectra of SN1998bw, where 
the OI] 6300 was narrower than the FeII] 5200 blend. \cite{mazzali2001,maeda2002}

Along the $z$-axis, heavy isotopes which are produced with high temperatures 
$T_9 \equiv T/10^9$K $\gsim 5$ are blown up to the surface.  
As a result, $^{64}$Ge, $^{59}$Cu, $^{56}$Ni, $^{48}$Cr, and $^{44}$Ti 
(which decay into $^{64}$Zn, $^{59}$Co, $^{56}$Fe, $^{48}$Ti, and $^{44}$Ca, respectively) 
are ejected at the highest velocities. 
Isotopes which are synthesized with somewhat lower temperatures ($T_9 = 4 - 5$) 
are first pushed aside as the jets propagate, then experience circulation to flow 
into behind the working surface. 
$^{55}$Co, $^{52}$Fe (which decay into $^{55}$Mn and $^{52}$Cr, respectively), 
$^{40}$Ca, $^{32}$S, and $^{28}$Si are 
therefore ejected at the intermediate velocities. 
Isotopes which are not synthesized but are only 
consumed during the explosion are accreted from the side. 
$^{24}$Mg, $^{16}$O, and $^{12}$C occupy the innermost region at the lowest velocities. 

The distribution of isotopes as a function of the velocity shows inversion as compared 
with conventional spherical models. This affects the overall abundance patterns 
in the whole ejecta as shown in Figure 3. 
As noted above, materials which experience higher $T_9$ 
are preferentially ejected along the $z$-axis, 
while materials with lower $T_9$ accrete from the side in the bipolar models. 
Zn and Co are ejected at higher velocities than Mn and Cr, so that the latter 
accrete onto the central remnant more easily. 
As a consequence, [Zn/Fe] and [Co/Fe] ([X/Fe] $\equiv \log10 ({\rm X/Fe}) - \log10({\rm X/Fe})_{\odot}$) 
are enhanced, while [Mn/Fe] and [Cr/Fe] are suppressed. 

\begin{figure}[t]
	\begin{center}
		\begin{minipage}[t]{0.6\textwidth}
			\includegraphics[width=1.\textwidth]{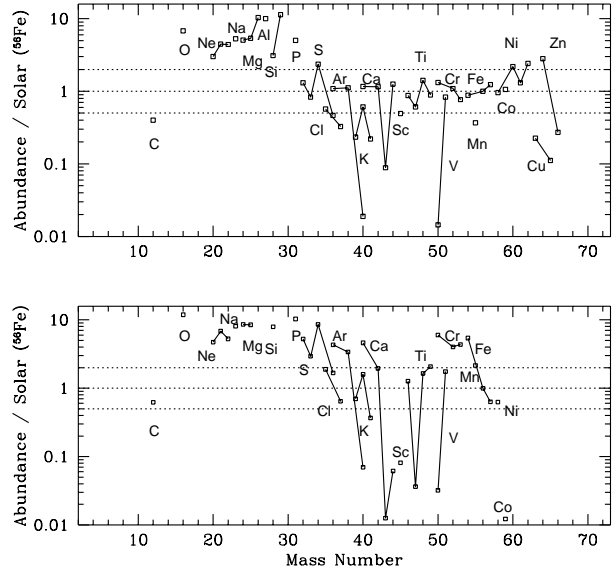}
		\end{minipage}
	\end{center}
\caption{Isotropic yields of the bipolar model 40A (upper) and a spherical model (lower) 
with $E_{51} = 10$, $M$($^{56}$Ni) $= 0.1\msun$, and $M_{\rm MS}=40\msun$. 
\label{fig:fig3}}
\end{figure}

Finally, we predict the relation among the optical luminosity, nucleosynthetic yield, 
and $M_{\rm REM}$ in the bipolar models. 
Figure 4 shows $M$($^{56}$Ni) as a function of $M_{\rm REM}$. 
For a given progenitor, they show a clear anti-correlation. 
As the optical luminosity is basically proportional to $M$($^{56}$Ni), 
we predict a smaller $M_{\rm REM}$ for a brighter supernova 
in the context of the bipolar explosion. 
The ratios among various elements depend on $M_{\rm REM}$. 
For example, [O/C] and [S/Si] are smaller for larger $M_{\rm REM}$ 
of the same progenitor.

\section{Conclusions}

We have presented hydrodynamic and nucleosynthetic features of the bipolar models. 
We have discussed advantages of our models in explaining the observed features of hypernovae, 
which conventional spherical models fail to reproduce. 

The bipolar models predict the following unique features. 
(1) Iron peak elements (e.g., Zn, Co, Fe) are blown up to the surface. 
The elements near the surface may be 
easily accelerated and ejected as cosmic rays. 
(2) (Zn, Co)/Fe are enhanced, while (Mn, Cr)/Fe are suppressed. 
These trends are seen in the abundances in extremely metal-poor stars, 
which likely lock up the information on 
chemical composition of supernovae at the earliest phase of the Galaxy. 
This might indicate a significant contribution of bipolar supernovae/hypernovae 
to the early Galactic chemical evolution. \cite{maeda2003b,maeda2003c,nomoto2003b}
(3) Certain relations among the optical luminosity, chemical composition, and $M_{\rm REM}$ 
are predicted. 
As $M_{\rm REM}$ is probably an indicator of gravitational emissions from the event, 
we expect the relation among the optical luminosity, chemical composition, and gravitational wave. 
Such relations may serve for identifying gravitational wave emitters in the future, 
through observations by different methods.

\begin{figure}[t]
\begin{center}
	\begin{minipage}[t]{0.5\textwidth}
			\includegraphics[width=0.925\textwidth]{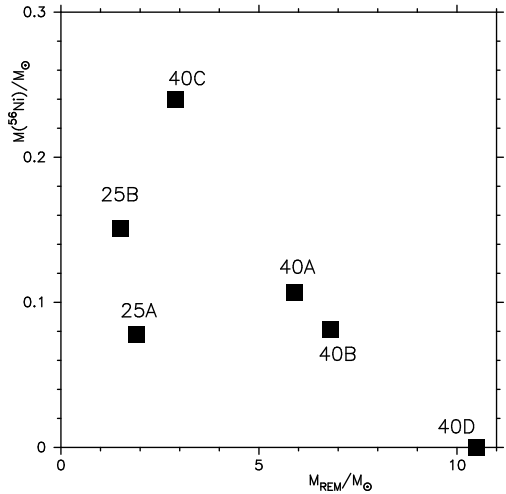}
	\end{minipage}
\end{center}
\caption{The mass of ejected $^{56}$Ni as a function of $M_{\rm REM}$. 
\label{fig:fig4}}
\end{figure}


%

\end{document}